Original Paper

# "Thought I'd Share First" and Other Conspiracy Theory Tweets from the COVID-19 Infodemic: Exploratory Study

Dax Gerts[1*], MS; Courtney D Shelley[1*], PhD; Nidhi Parikh[1], PhD; Travis Pitts[1], MA; Chrysm Watson Ross[1,2], MA, MS; Geoffrey Fairchild[1], PhD; Nidia Yadria Vaquera Chavez[1,2], MS; Ashlynn R Daughton[1], MPH, PhD

[1]Analytics, Intelligence, and Technology Division, Los Alamos National Laboratory, Los Alamos, NM, United States

[2]Department of Computer Science, University of New Mexico, Albuquerque, NM, United States

[*]these authors contributed equally

**Corresponding Author:**
Ashlynn R Daughton, MPH, PhD
Analytics, Intelligence, and Technology Division
Los Alamos National Laboratory
P.O. Box 1663
Los Alamos, NM, 87545
United States
Phone: 1 505 664 0062
Email: adaughton@lanl.gov

## Abstract

**Background:** The COVID-19 outbreak has left many people isolated within their homes; these people are turning to social media for news and social connection, which leaves them vulnerable to believing and sharing misinformation. Health-related misinformation threatens adherence to public health messaging, and monitoring its spread on social media is critical to understanding the evolution of ideas that have potentially negative public health impacts.

**Objective:** The aim of this study is to use Twitter data to explore methods to characterize and classify four COVID-19 conspiracy theories and to provide context for each of these conspiracy theories through the first 5 months of the pandemic.

**Methods:** We began with a corpus of COVID-19 tweets (approximately 120 million) spanning late January to early May 2020. We first filtered tweets using regular expressions (n=1.8 million) and used random forest classification models to identify tweets related to four conspiracy theories. Our classified data sets were then used in downstream sentiment analysis and dynamic topic modeling to characterize the linguistic features of COVID-19 conspiracy theories as they evolve over time.

**Results:** Analysis using model-labeled data was beneficial for increasing the proportion of data matching misinformation indicators. Random forest classifier metrics varied across the four conspiracy theories considered (F1 scores between 0.347 and 0.857); this performance increased as the given conspiracy theory was more narrowly defined. We showed that misinformation tweets demonstrate more negative sentiment when compared to nonmisinformation tweets and that theories evolve over time, incorporating details from unrelated conspiracy theories as well as real-world events.

**Conclusions:** Although we focus here on health-related misinformation, this combination of approaches is not specific to public health and is valuable for characterizing misinformation in general, which is an important first step in creating targeted messaging to counteract its spread. Initial messaging should aim to preempt generalized misinformation before it becomes widespread, while later messaging will need to target evolving conspiracy theories and the new facets of each as they become incorporated.

## Introduction

### Background

On December 31, 2019, the World Health Organization (WHO) was made aware of a cluster of cases of viral pneumonia of unknown origin in Wuhan, Hubei Province, China [1]. The WHO reported this cluster via Twitter on January 4, 2020, saying, "#China has reported to WHO a cluster of #pneumonia cases —with no deaths— in Wuhan, Hubei Province. Investigations are underway to identify the cause of this illness [2]." On January 19, the WHO Western Pacific Regional Office tweeted evidence of human-to-human transmission, saying, "According to the latest information received and @WHO analysis, there is evidence of limited human-to-human transmission of #nCOV. This is in line with experience with other respiratory illnesses and in particular with other coronavirus outbreaks [3]." The first case in the United States was reported the next day. Five days later, on January 26, 2020, *GreatGameIndia* published the article "Coronavirus Bioweapon–How China Stole Coronavirus From Canada And Weaponized It," which claimed that the coronavirus was leaked into China from a Canadian laboratory [4]. The original article received 1600 likes on its first day of publication; it was then reposted verbatim but with the more provocative headline "Did China Steal Coronavirus From Canada And Weaponize It" on the website *ZeroHedge* [5]. This version was reposted by the website RedStateWatcher.com, one of the 140 most popular sites in the United States, with more than 4 million followers on Facebook; from there, the story quickly went viral [6].

Misinformation surrounding pandemics is not unique to SARS-CoV-2, the virus that causes COVID-19. At least as far back as the Russian flu pandemic of 1889, pandemic spread of *misinformation*, claims of fact that are either demonstrably false or unverifiable [7], has been concomitant with disease spread [8]. People are susceptible to misinformation when trust in authoritative sources is low, which can occur when officials provide conflicting information and guidance [9]. Misinformation will also include *conspiracy theories*, which posit explanations of events or circumstances based primarily on a conspiracy [10] (ie, an agreement between a small group of people to commit an illegal act). Although some conspiracies, such as Watergate or the Tuskegee experiments, may eventually be proven to be true criminal acts, the vast majority of conspiracy theories are not true, and their spread can undermine public health efforts [11]. Some conspiracy theories may be better classified as *disinformation*—false or misleading information that is intentionally passed to a target group [12] with its true source concealed [13].

The COVID-19 outbreak has left many people isolated within their homes, and these people are turning to social media for news and social connection. Thus, they are especially vulnerable to believing and sharing conspiracy theories [14]. This study examines four oft-repeated and long-lived conspiracy theories surrounding COVID-19: 5G technology is somehow associated with the disease; Bill Gates or the Bill & Melinda Gates Foundation created or patented the virus; the virus is human-made and was released from a laboratory; and a COVID-19 vaccine will be harmful. None of these conspiracy theories are unique, nor are they entirely distinct.

### 5G Cell Towers Spread COVID-19

Cellular carriers began a limited rollout of 5G cellular service in 2018 [15], which required the installation of new cell towers [16]. These new towers were already the source of a more general conspiracy theory that the signal is harmful to humans and that its dangers were being "covered up" by "powerful forces in the telecommunications industry" [17]. Wireless technology has consistently been blamed for causing immune damage in humans, and similar theories were seen with the rollouts of 2G, 3G, 4G, and Wi-Fi service [17]. Even the 1889 Russian flu was purported to be caused by the then-new technology of electric light [8]. The COVID-19–related 5G conspiracy theory emerged in the first week of January, and it may not have evolved past a fringe view into a trending hashtag without being shared by websites with the primary aim of spreading conspiracy theories on Twitter or by people aiming to denounce the theory [18].

### Bill Gates and the Bill & Melinda Gates Foundation

J Uscinski stated that conspiracy theories often "are about accusing powerful people of doing terrible things" [19]. The Bill & Melinda Gates Foundation is arguably the largest philanthropic venture ever attempted, and it has proven to be fertile ground for the development of conspiracy theories, ranging from misinterpretations of a "patent on COVID-19" [20] to incorporation of vaccine-averse concerns. For example, the Bill & Melinda Gates Foundation funded research to develop injectable invisible ink to serve as a permanent record of vaccination in developing countries [21,22]. This technology was announced in December 2019, the same month that SARS-CoV-2 emerged in Wuhan, China, and a conspiracy theory emerged suggesting that the COVID-19 vaccine would be used to microchip individuals with the goal of population control [20].

### Laboratory Origins

Associations between HIV and other infectious diseases consistently re-emerge, including associations with polio [23], Ebola virus [24], and COVID-19. The COVID-19–related HIV conspiracy theory began on January 31, 2020, with the preprint publication of "Uncanny similarity of unique inserts in the 2019-nCoV spike protein to HIV-1 gp120 and Gag" ([25], withdrawn paper), which was quickly retweeted by Anand Ranganathan, a molecular biologist with over 200,000 followers on Twitter. He cited the preprint as evidence of a potential laboratory origin with a now-deleted Tweet: "Oh my god. Indian scientists have just found HIV (AIDS) virus-like insertions in the 2019-nCoV virus that are not found in any other coronavirus. They hint at the possibility that this Chinese virus was designed…" Within two hours, Ross Douthat, a prominent *New York Times* opinion columnist, retweeted Ranganathan to his >140,000 followers, further legitimizing the theory through a reputable news outlet and greatly furthering the reach of the story outside the scientific community [26]. Three days after the initial release of the preprint, the original paper was retracted.

Laboratory origin theories have also garnered political attention; then-US President Donald Trump claimed to have evidence of a Chinese laboratory origin of SARS-CoV-2 [27], prompting a Twitter response from a Chinese government account [28] that was flagged by Twitter as misinformation [29]. Additional laboratory-related conspiracy theories quickly emerged, including theories that the virus was created to achieve global population reduction or to impose quarantines, travel bans, and martial law, all of which were previously seen during the 2014 Ebola virus outbreak [24] and the 2015-2016 Zika virus outbreak [30].

### Vaccines

Vaccine-related social media articles are often shared by people who are relatively knowledge-deficient and vaccine-averse compared to nonsharers [31], with content consisting of debunked associations with autism and general mistrust of government or the pharmaceutical industry. With newly emergent diseases such as HIV and Ebola, conspiracy theories quickly followed regarding the ability to profit off of vaccines while conspiring with American pharmaceutical companies [24].

In the past year, substantial work has emerged investigating the onslaught of misinformation related to COVID-19. Multiple studies have found that misinformation is common; both social media platforms [32-34] and web pages returned results for common COVID-19 queries at the beginning of the pandemic [35], including scientific journals without sufficiently rigorous review processes [36].

Social media studies have so far indicated that original tweets present false information more often than evidence-based information, but that evidence-based information is more often retweeted [32]; therefore, during the first three months of the outbreak, the volume of misinformation tweets was small compared to that of the overall conversation [37]. The amount of Twitter data related to COVID-19 dwarfed that of other health-related content, but proportionally more of the data originated from credible websites [33].

Researchers have also attempted to characterize the people who are likely to believe misinformation. One nationally representative study in the United States found that some myths (eg, that the virus was created or spread on purpose) were believed by over 30% of respondents [38]. Evidence across several countries shows that people who believe misinformation are more likely to obtain information from social media or have a self-perceived minority status [39], and characteristics such as “trusting scientists” and obtaining information from the WHO had a negative relationship with belief in misinformation [40].

With the above framing in mind, this paper seeks to answer the following research questions:

1. *Can conspiracy theories identified a priori be automatically identified using supervised learning techniques?*

We used a large corpus of Twitter data (120 million initial tweets and 1.8 million tweets after our initial regular expression filtering step) and random forest models to classify tweets associated with the four conspiracy theories described above.

2. *Can identified tweets about defined conspiracy theories be characterized by existing methodologies?*

We used tweet sentiment to assess the emotional valence in conspiracy theory tweets compared to their non–conspiracy theory counterparts. We used dynamic topic modeling, an unsupervised learning approach, to explore the changes in word importance among the topics within each theory.

3. *Can our findings inform public health messaging to reduce the effects of misinformation found on social media?*

We compared the results of the preceding research questions to identify commonalities and connections between early conspiracy theories that can be addressed by initial public health messaging to prevent further misinformation spread. We additionally showed that theories evolve to include real-world events and incorporate details from unrelated conspiracy theories; therefore, later public health messaging will also need to evolve.

# Methods

## Data

### *Twitter Data*

The Twitter data used for this study were derived from Chen et al (2020) [41], who constructed the tweet IDs of tweets that include COVID-19 keywords and health-related Twitter accounts and made them publicly available. Due to limitations in the Twitter application programming interface (API), these data represent a 1% sample of tweets that included these keywords or tracked accounts. We gathered these data from the Twitter API using the released IDs, identifying approximately 120 million tweets from January 21 to May 8, 2020 (see Figure 1). Although the initial repository includes tweets in a variety of languages [41], we restricted our analysis to tweets in English.

**Figure 1.** Volume of Twitter data collected during the study period. Twitter data were collected from January 21 to May 8, 2020, representing the first five months of the COVID-19 pandemic. We have annotated this timeline with major events to provide context during this early period of the pandemic. AZ: Arizona; NC: North Carolina; NH: New Hampshire; NY: New York; OR: Oregon; Trump: US President Donald Trump; US: United States; WHO: World Health Organization.

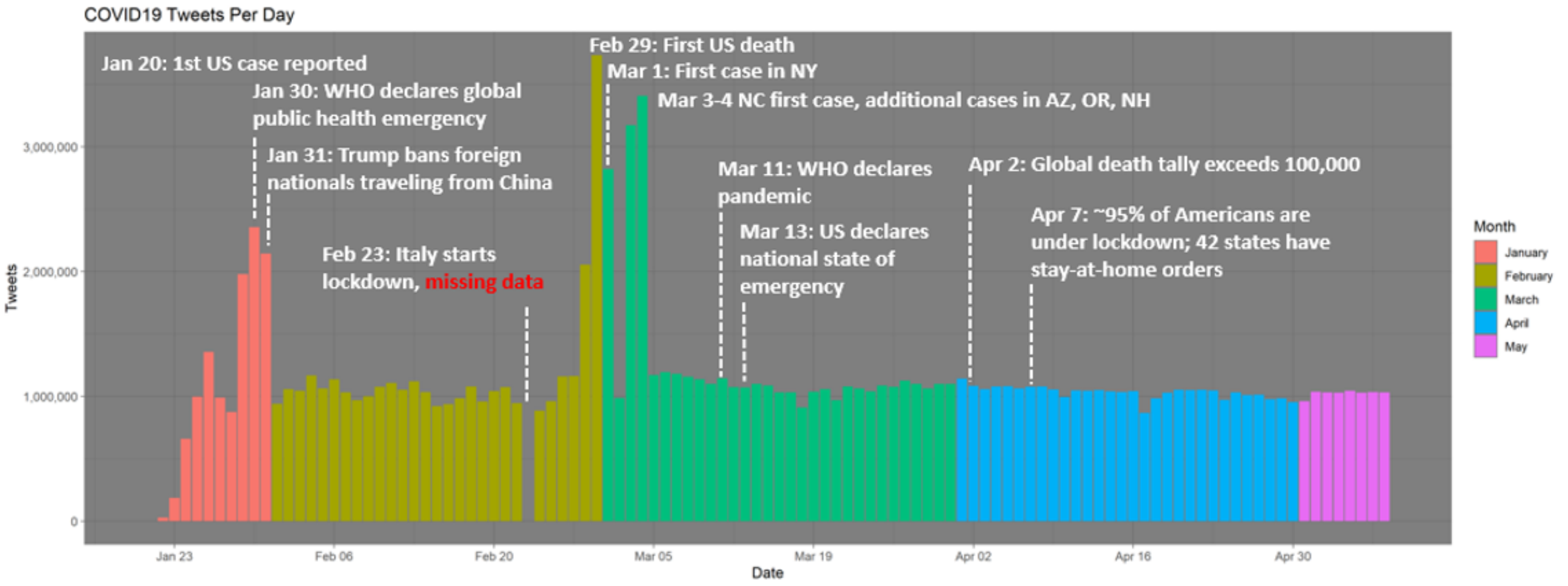

## *NewsGuard*

NewsGuard provides evaluations of thousands of websites based on criteria including funding transparency, journalistic integrity, and editorial track record [42]. Since the emergence of COVID-19, NewsGuard has also provided a summary of major myths and conspiracy theories associated with the pandemic, the earliest documented claims, major events that caused significant spread, and detailed reports of major sources of COVID-19 misinformation in their "Special Report: COVID-19 Myths" [20]. From this list, we identified four theories that were especially prominent in our Twitter data set and that were commonly discussed in mainstream news media. In addition, we used the domains classified as "not credible" and related to COVID-19 myths, as identified by NewsGuard, as features in our classification models described below.

## Filtering and Supervised Classification

We filtered the data into four data sets using regular expressions (see Figure 2) to increase the number of relevant tweets in each category of interest [43-47]. The four data sets are hereafter referred to using the following terms:

- 5G: 5G technology is somehow associated with COVID-19.
- Gates: Bill and Melinda Gates or the Bill & Melinda Gates Foundation funded, patented, or otherwise economically benefited from SARS-CoV-2.
- Lab: SARS-CoV-2 is human-made or bioengineered and was released (intentionally or accidentally) from a laboratory.
- Vax: A COVID-19 vaccine would be harmful in a way not supported by science (eg, it could contain a microchip).

**Figure 2.** Tweet-filtering flow. The initial tweet corpus was obtained from Chen et al [41], who used keywords and known accounts to provide a sample of COVID-19–related Twitter data (Filter 1). We then used regular expressions to create four conspiracy theory data sets (Filter 2) and machine learning classifiers to identify misinformation tweets within each data set (Filter 3). 5G: conspiracy theories related to 5G technology; CDC: US Centers for Disease Control and Prevention; Gates: conspiracy theories related to Bill Gates or the Bill & Melinda Gates Foundation, Lab: conspiracy theories related to the virus being laboratory-released or human-made; Vax: conspiracy theories related to vaccines; WHO: World Health Organization.

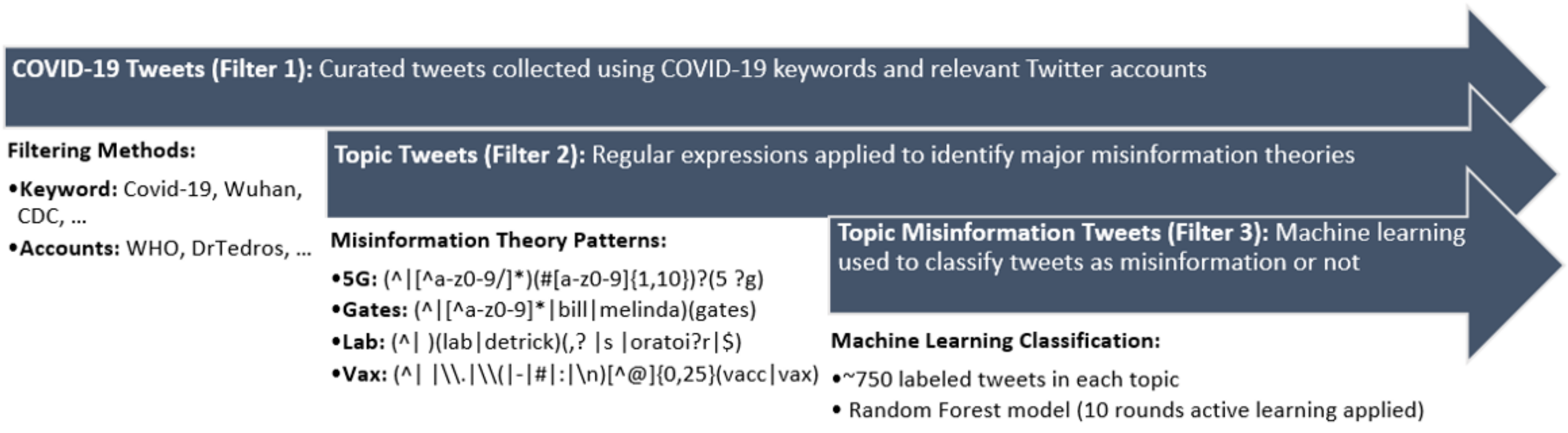

Within each regular expression-filtered conspiracy theory data set, we randomly sampled 1000 tweets to create the training data. After sampling, duplicate tweets were removed. Two authors coded each set of tweets and established agreement by jointly coding a subset of tweets (see Table 1). Any tweet promoting or engaging with misinformation, even to refute it, was labeled as COVID-19 misinformation. This labeling was performed with the rationale that tweeting about misinformation, even in the context of a correction, increases the size of the audience exposed to that misinformation. In prior work on COVID-19 conspiracy theories, it was found that engaging with a theory to correct it can indeed increase the overall visibility of the theory [18]. Interrater analysis found relatively high agreement and reasonable Cohen κ scores (mean 0.759, Table 1). However, the effort demonstrated the difficulty of reliably identifying misinformation; in many cases, oblique references and jokes fell in a gray area that raters labeled "uncertain" (~6.1% of the coded tweets). A second pass was made over tweets labeled "uncertain" by comparing rater assessments and marking these tweets as "COVID-19 misinformation" or "not COVID-19 misinformation" based on rater agreement. For example, if annotators 1 and 2 had high agreement when labeling 5G tweets, a tweet labeled by annotator 1 as "uncertain" could be relabeled as "COVID-19 misinformation". Using this approach, we were able to avoid removing data and thus shrinking the amount of available training data.

**Table 1.** Interrater results from the creation of the training data. Tweets were randomly sampled from the regular expression-filtered data sets and duplicates were removed. Each rater was assigned a portion of overlapping tweets to allow for interrater evaluation.

| Theory | Unique tweets labeled (n) | Tweets labeled by multiple authors (n) | Agreement | Cohen κ |
|---|---|---|---|---|
| 5G[a] | 725 | 146 | 0.852 | 0.708 |
| Gates[b] | 711 | 143 | 0.893 | 0.782 |
| Lab[c] | 735 | 146 | 0.901 | 0.796 |
| Vax[d] | 775 | 199 | 0.915 | 0.751 |

[a]5G: conspiracy theories related to 5G technology.

[b]Gates: conspiracy theories related to Bill Gates or the Bill & Melinda Gates Foundation.

[c]Lab: conspiracy theories related to SARS-CoV-2 being laboratory-released or human-made.

[d]Vax: conspiracy theories related to vaccines.

The tweets were tokenized, and both URLs and stop words were removed. Unigrams and bigrams were used as features in a document-term matrix, and the most sparse (<0.05% populated) terms were removed. Additionally, we added Boolean features describing relationships to domains identified by NewsGuard as sources of misinformation. This was achieved by linking associated Twitter accounts to tracked websites. Features included (1) a tweet originating from a misinformation-identified domain, (2) a tweet replying to an originating tweet, (3) a tweet retweeting an originating tweet, or (4) a tweet that was otherwise linked (eg, replying to a retweet of a tweet from a misinformation source). As noted elsewhere, only English tweets were used in this analysis.

The data were partitioned into a two-thirds/one-third training-test split. Data were sampled so that the training data had an equal sample distribution (50% misinformation, 50% nonmisinformation). The testing data used the remaining available data; thus, the sample distribution was uneven.

Classifiers were built using R, version 3.6.3 (R Project); the randomForest package, version 4.6-14, was used to train random forest models with 150 trees up to 25 terminal nodes (and at least 3 terminal nodes), and 25 variables were randomly sampled at each split. Case sampling was performed with replacement. We used an active learning approach in which after each run of the random forest classifier, the calculated posterior entropy was used to select the three unlabeled tweets that caused the most uncertainty in the model. These were then hand-labeled by an author (DG) and applied to the next run of the model. We applied 9 cycles of active learning to each model. Additionally, for each hand-labeled tweet, highly similar tweets (string similarity ≥0.95) were identified and given the same label. This approach was implemented using the R activelearning package, version 0.1.2. The models that performed the best (measured by F1 score) were used to assign labels to the regular expression-filtered tweets.

## Sentiment Analysis

Two well-documented sentiment dictionaries were used to label the tokenized tweets. The first, AFINN [48], provided an integer score ranging from –5 (negative sentiment) to +5 (positive sentiment) for each word in the dictionary. The second dictionary, the National Research Council (NRC) Word-Emotion Association Lexicon [49], was used to tag words with categories of emotion, providing labels for 8 emotions of anger, anticipation, disgust, fear, joy, sadness, surprise, and trust in addition to an overall "positive" or "negative" sentiment. We then compared the sentiment for each classified data set over time. For each tweet, aggregate sentiment metrics were calculated, including the sum of integer scores and the counts for each emotion label.

## Dynamic Topic Modeling

Dynamic topic modeling (DTM) was used to characterize themes and analyze temporal changes in word importance [50]. DTM divides tweets into weekly time slices based on the time they were generated. The set of topics at each time slice is then assumed to evolve from the set of topics at the previous time slice using a state space model. The result is an evolving probability distribution of words for each topic that shows how certain words become more or less important over time for the same topic. Traditional topic models, such as latent Dirichlet allocation [51], assume that all the documents (which are here equivalent to tweets) are drawn exchangeably from the same topic distribution, irrespective of the time when they were generated. However, a set of documents generated at different times may reflect evolving topics.

Dynamic topic models were trained for each conspiracy theory, with the number of topics ranging from 2-5. Small numbers of topics were chosen because these tweets were already classified to be relevant for individual misinformation topics, and because our goal was to identify potential subtopics that evolved over time. The optimal number of topics was assessed qualitatively by reviewing the topic modeling results. DTM was implemented in Python using the gensim [52] wrapper ("ldaseqmodel") for the DTM model [50,53].

## Results

### Filtering and Supervised Classification

After filtering using regular expressions, our corpus included roughly 1.8 million unique tweets across the four conspiracy theories (Table 2). The relative volume of tweets in each data set is shown in Figure 3. The number of tweets appearing in multiple data sets corresponds to the edge thickness. All the data sets showed some degree of overlap between categories, with Gates showing the most overlap and 5G showing the least. 5G additionally had a low volume of tweets compared to the other theories.

**Table 2.** Results of the regular expression filtering step. After filtering using regular expressions on tweets spanning January 21 to May 8, 2020, the number of tweets per conspiracy theory and the number of tweets that were included in multiple theories are shown. The number of tweets within each filtered data set that were later classified as COVID-19 misinformation and the number of classified tweets that appear in multiple theories are also provided.

| Conspiracy theory | Tweets after regular expression filtering (n=1,901,108), n (%) | Tweets after regular expression filtering found in multiple theories, n (%) | Tweets classified as COVID-19 misinformation, n (%) | Tweets classified as COVID-19 misinformation found in multiple theories, n (%) |
|---|---|---|---|---|
| 5G[a] | 127,209 (6.69) | 6300 (4.95) | 51,049 (40.13) | 1984 (1.56) |
| Gates[b] | 278,130 (14.63) | 69,566 (25.01) | 147,657 (53.09) | 35,880 (12.90) |
| Lab[c] | 526,115 (27.64) | 44,198 (8.40) | 224,052 (42.59) | 20,001 (3.80) |
| Vax[d] | 969,654 (51.00) | 82,380 (8.50) | 206,046 (21.25) | 34,435 (3.55) |

[a]5G: conspiracy theories related to 5G technology.

[b]Gates: conspiracy theories related to Bill Gates or the Bill & Melinda Gates Foundation.

[c]Lab: conspiracy theories related to SARS-CoV-2 being laboratory-released or human-made.

[d]Vax: conspiracy theories related to vaccines.

**Figure 3.** Data set volumes and overlap by theory. The node size indicates the total number of tweets discussing each conspiracy theory, while the edge thickness corresponds to the number of tweets discussing any pair of conspiracy theories simultaneously. 5G: conspiracy theories related to 5G technology; Gates: conspiracy theories related to Bill Gates or the Bill & Melinda Gates Foundation; Lab: conspiracy theories related to SARS-CoV-2 being laboratory-released or human-made; Vax: conspiracy theories related to vaccines.

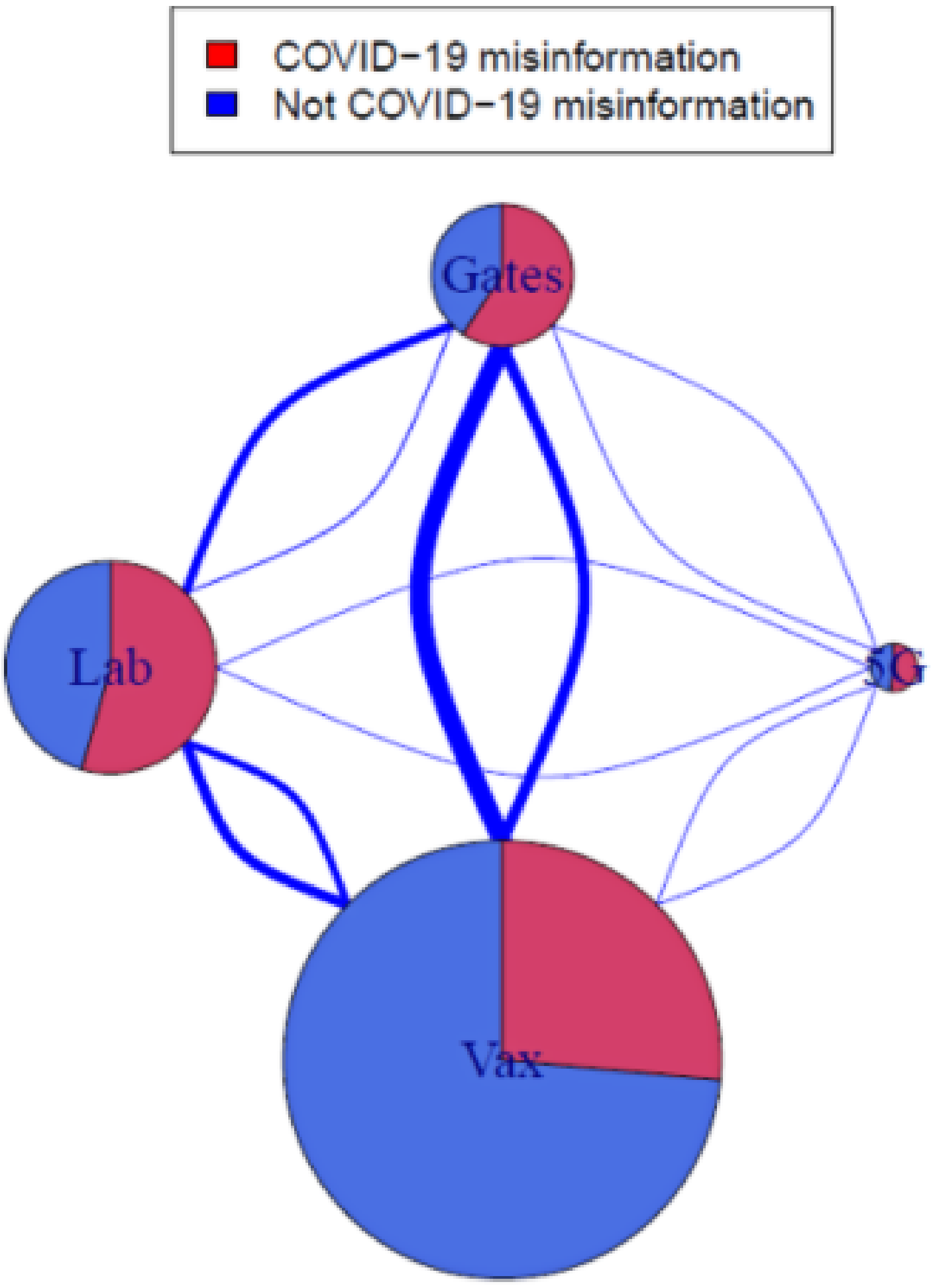

The model performance metrics for each theory are presented in Tables 3 and 4. Class proportions were roughly balanced in the 5G, Gates, and Lab theories. The Vax tweets were heavily imbalanced, with only ~18% labeled as COVID-19 misinformation (Table 3). The best performing models were the 5G and Lab theories, with F1 scores of 0.804 and 0.857, respectively (Table 4). Although the results for the Gates theory were weaker (F1 score=0.654), Vax scored the lowest (F1 score=0.347). This could be due to the imbalanced nature of the data set.

**Table 3.** Distributions of labels for the four COVID-19 conspiracy theories.

| Conspiracy theory | Label distribution | | |
|---|---|---|---|
| | COVID-19 misinformation, n | Not COVID-19 misinformation, n | Proportion of COVID-19 misinformation, % |
| 5G[a] | 367 | 356 | 50.8 |
| Gates[b] | 354 | 356 | 49.9 |
| Lab[c] | 407 | 327 | 55.4 |
| Vax[d] | 142 | 632 | 18.3 |

[a]5G: conspiracy theories related to 5G technology.

[b]Gates: conspiracy theories related to Bill Gates or the Bill & Melinda Gates Foundation.

[c]Lab: conspiracy theories related to SARS-CoV-2 being laboratory-released or human-made.

[d]Vax: conspiracy theories related to vaccines.

**Table 4.** Random forest model results. Random forest with active learning often, although not universally, shows improved performance compared to generic random forest models. The change between these two approaches is noted in the Change column.

| Conspiracy theory and metrics | Random forest | Random forest with active learning | Change |
|---|---|---|---|
| **5G[a]** | | | |
| Accuracy | 0.779 | 0.783 | 0.004 |
| Recall | 0.908 | 0.872 | –0.036 |
| Precision | 0.728 | 0.744 | 0.016 |
| F1 Score | 0.808 | 0.804 | –0.004 |
| **Gates[b]** | | | |
| Accuracy | 0.622 | 0.5819 | –0.04 |
| Recall | 0.675 | 0.793 | 0.118 |
| Precision | 0.608 | 0.556 | -0.052 |
| F1 Score | 0.64 | 0.654 | 0.014 |
| **Lab[c]** | | | |
| Accuracy | 0.782 | 0.84 | 0.058 |
| Recall | 0.699 | 0.833 | 0.134 |
| Precision | 0.9 | 0.883 | –0.017 |
| F1 Score | 0.787 | 0.857 | 0.070 |
| **Vax[d]** | | | |
| Accuracy | 0.507 | 0.751 | 0.244 |
| Recall | 0.653 | 0.474 | –0.1786 |
| Precision | 0.170 | 0.274 | 0.104 |
| F1 Score | 0.270 | 0.347 | 0.077 |

[a]5G: conspiracy theories related to 5G technology.

[b]Gates: conspiracy theories related to Bill Gates or the Bill & Melinda Gates Foundation.

[c]Lab: conspiracy theories related to SARS-CoV-2 being laboratory-released or human-made.

[d]Vax: conspiracy theories related to vaccines.

## Sentiment Analysis

The range in sentiment was significantly greater for COVID-19 misinformation, with tweets more consistently showing increased negative sentiment, especially in April and May 2020. Figure 4 shows Gates-related tweets by net sentiment score over time. See Multimedia Appendix 1 for additional figures related to other conspiracy theories (Figures S1-S3).

**Figure 4.** Sentiment comparison for data from tweets about COVID-19 conspiracy theories related to Bill Gates and the Bill & Melinda Gates Foundation by label. Tweets are plotted over time and stratified by misinformation status. Sentiment varies from highly negative to highly positive. Loess smoothing was used to draw the blue line indicating general trend over time.

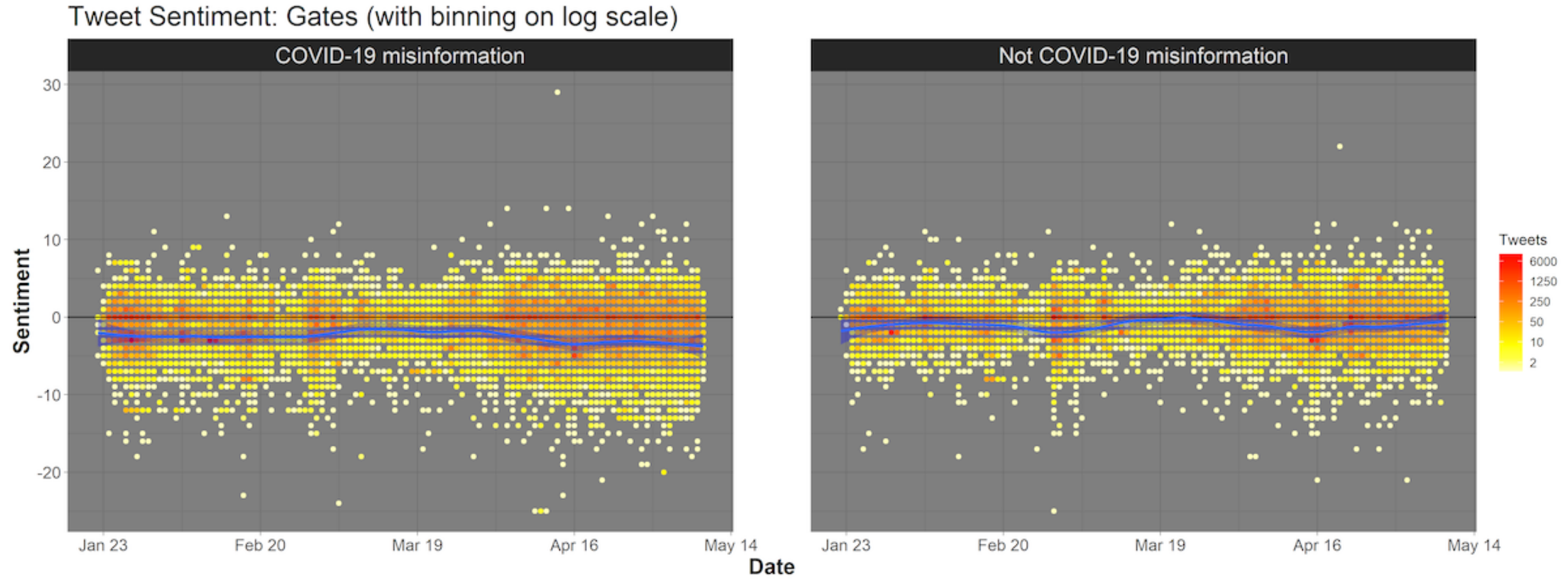

Figure 5 shows the sentiments of tweets (with daily average sentiment scores for each category averaged across all dates in the study range) in each conspiracy theory subset across eight emotions and the general negative or positive sentiment. Although tweets related to 5G conspiracies show similar results for misinformation and nonmisinformation, there are clear differences in the other four conspiracy theories. In general, tweets classified as misinformation tend to rate higher on negative sentiment, fear, anger, and disgust compared to tweets not classified as misinformation.

**Figure 5.** Sentiment comparison for each conspiracy theory by classification. The average numbers of words per tweet flagged for each sentiment category are plotted. 5G: conspiracy theories related to 5G technology; Gates: conspiracy theories related to Bill Gates or the Bill & Melinda Gates Foundation; Lab: conspiracy theories related to SARS-CoV-2 being laboratory-released or human-made; Vax: conspiracy theories related to vaccines.

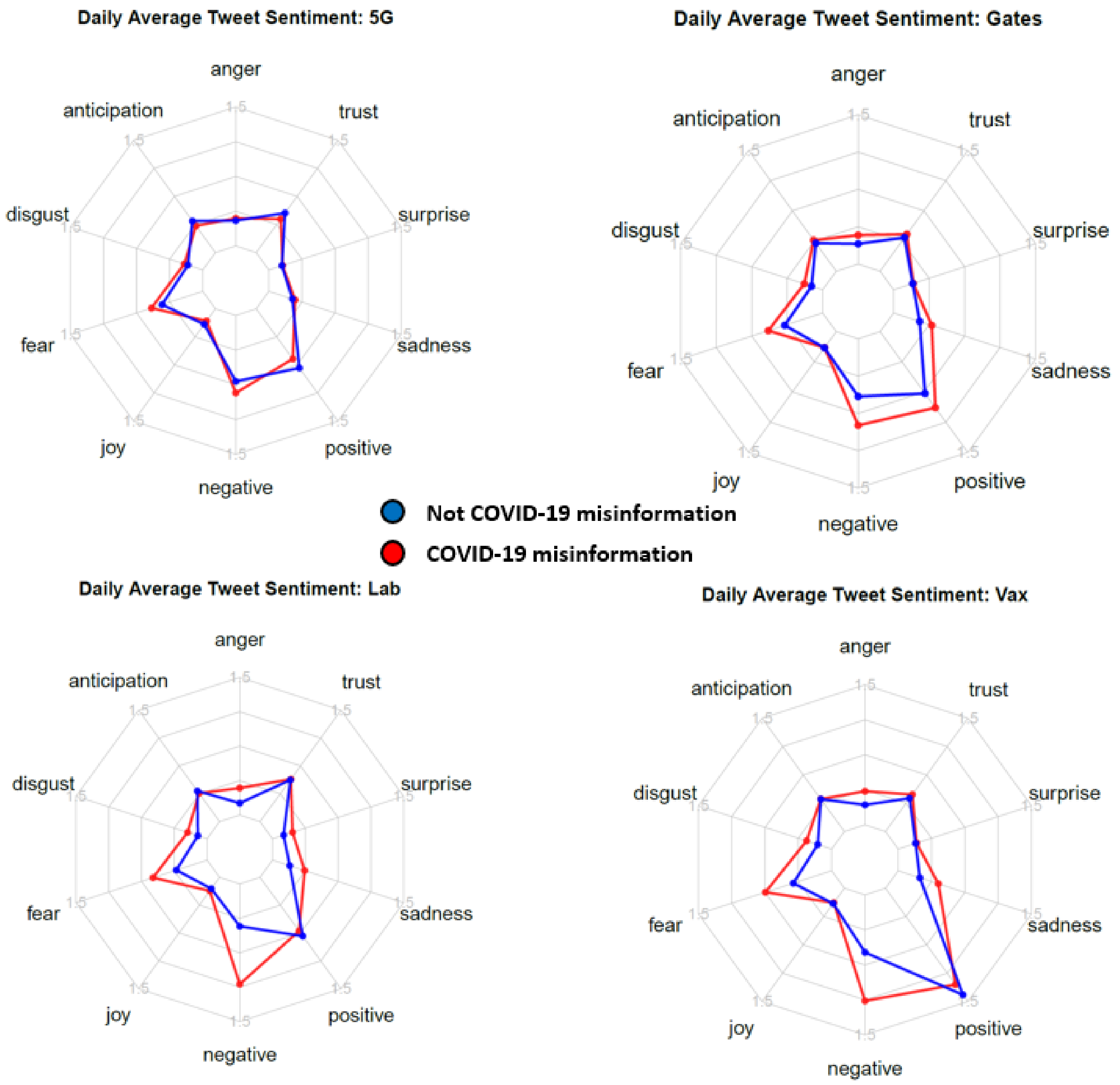

## DTM Analysis

For each conspiracy theory data set, DTM was used to identify 2-5 potential subtopics and understand their evolution over time. The optimal model was assessed qualitatively by reviewing the results. Models with 2 topics led to optimal results (qualitatively coherent topics with the least amount of overlap) for Gates, 5G, and Lab theories, while the model with 3 topics qualitatively led to optimal results for Vax theories. The results for the Gates theory are visualized here, and the remaining theories are visualized in Multimedia Appendix 1.

The Gates theory was optimally represented by 2 topics. Both topics showed peaks of increased Twitter discussion in mid-January to mid-February and a second peak in April (Figure 6). The initial peaks in Topic 1 corresponded to high weighting of the words *predicted*, *kill_65m*, *event*, and *simulation*, while the later spike in April showed higher weights for words such as *fauci* and *buttar* (Figure 7). The model identified a second topic that referred to several conspiracy theories about Bill Gates, SARS-CoV-2, and vaccines. This second topic initially focused on theories about the origins of the virus, with highly weighted words including *pirbright* and *patent*. In late April, higher-weighted words included *kennedy*, *jr*, and *fauci*.

The Vax data showed high weighting for the word *bakker* in Topic 1 and a brief increase in the word *microchip* in early April within Topic 2 (Multimedia Appendix 1, Figure S6). The term *bakker* refers to the tele-evangelist Jim Bakker, who promoted myths about possible COVID-19 cures, including the use of colloidal silver, on his show [54]. A linguistic shift in referring to the virus was also observable within the vaccine theory, with *coronavirus* highly weighted until mid-March, when *COVID* became more frequently used.

In the Lab data, words such as *biosafety*, *biowarfare*, *warned*, and *laboratory* were more highly weighted early in the outbreak, suggesting that people were discussing a malicious laboratory release [63] (Multimedia Appendix 1, Figure S5, topic 2*)*. The weight of words such as *escaped*, *evidence*, and *originated* increased as the theory evolved over time. Overlap was seen between the Lab theory and the Gates theory, including words such as *kill*, *kill_65m*, and *kill_forget*. In addition, we observed terms related to other, older theories, such as *ebola* in Topic 2 in mid-January, and terms related to Jeffrey Epstein and conspiracy theories associated with his death (*epstein*, *forget_epstein*) [40].

**Figure 6.** Topic distribution over time for the 2-topic dynamic topic model for tweets related to the conspiracy topic of Bill Gates and the Bill & Melinda Gates Foundation. Tweets belonging to Topic 1 are more common in the conversation in January, while Topic 2 becomes more prominent in the spring. Additionally, distinct peaks show the popularity of tweets related to this conspiracy theory category overall.

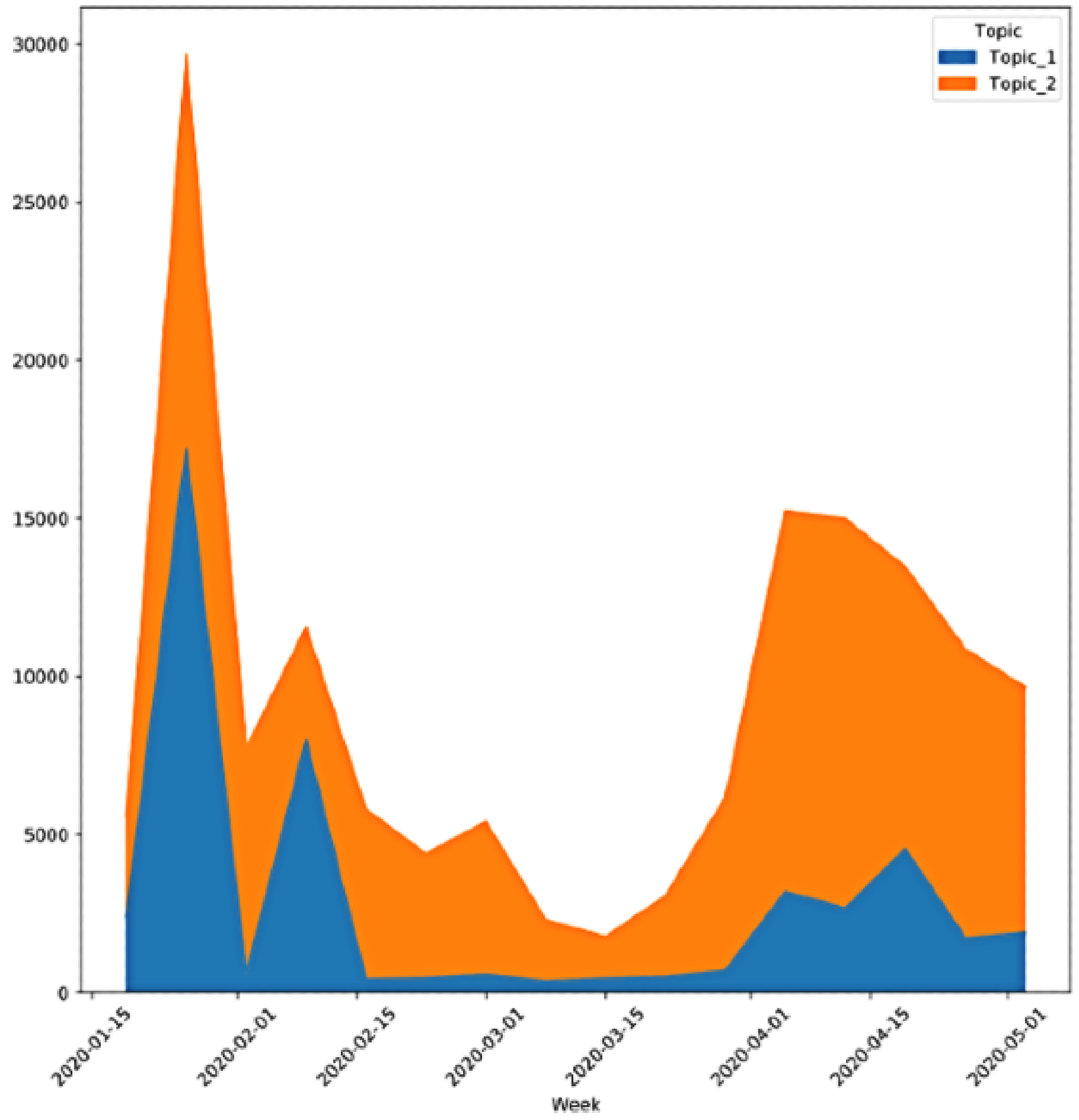

**Figure 7.** Topic evolutions and word clouds for COVID-19 conspiracy theories related to Bill Gates and the Bill & Melinda Gates Foundation. Top panel: word evolution in the 2-topic dynamic topic model. Color represents the importance of the words, with a darker color denoting higher importance. Bottom panel: word clouds for each topic. The size of each word corresponds to its weight (higher-weighted words are larger in size).

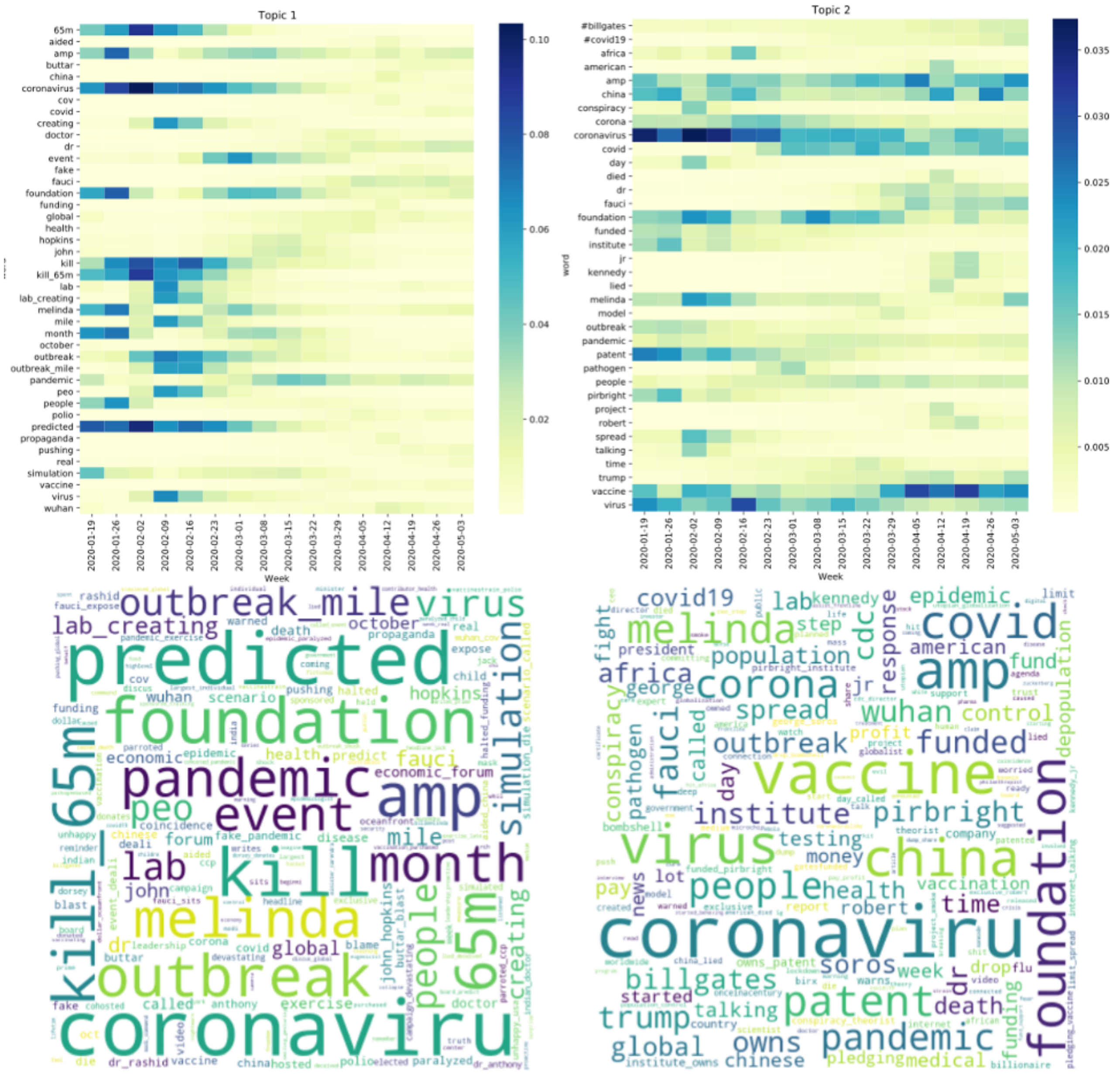

## Discussion

### Principal Findings

The ongoing COVID-19 pandemic clearly illustrates the need to identify health-related misinformation, especially with a lens toward improving communication strategies to combat it. We focused on four specific conspiracy theories and fused existing methods to identify relevant tweets and characterize the language used over time. This is especially important in the context of COVID-19 as an emerging infectious disease, when much of the scientific knowledge about its risks, transmission, and mitigation may be quickly evolving [56]. With this context in mind, we address our findings with respect to each research question below.

### *Can Conspiracy Theories Identified A Priori Be Automatically Identified Using Supervised Learning Techniques?*

In prior work, it was found that misinformation, defined more broadly than just conspiracy theories, is relatively common on social media [32-34], with some caveats. For example, although original tweets were found to present false information more often than evidence-based information, evidence-based information was retweeted more often [32]. In another analysis, it was found that although a greater proportion of data on Twitter originated from credible websites than from noncredible websites, there were instances in which low-quality content was boosted by credible websites; it was also found that website credibility may be a poor marker of the quality of information being presented [33]. Overall, we classified 582,290 tweets (32% of the regular expression–filtered corpus) as relating to at least one of the four specific conspiracy theories considered. Using regular expression–based filtering and supervised learning, we identified tweets associated with these conspiracy theories. Classification models performed quite well for the 5G and Lab theories because the focus of these conspiracy theories was well defined. Classifiers for the Gates conspiracy theory performed more moderately, likely because the theory was broad and its content overlapped with that of the other three theories. The Vax theory performed the worst, likely due to class imbalance.

### *Can Identified Tweets About Defined Conspiracy Theories Be Characterized by Existing Methodologies?*

We used sentiment analysis to assess the affective states of tweets classified as misinformation. Overall, misinformation-classified tweets showed more negative sentiment over time, both on a scale from negative to positive sentiment and when discretized into specific emotions. Within specific conspiracy theories, these differences were the smallest when comparing misinformation and nonmisinformation in the 5G data. This could be a result of the intense political polarization surrounding the rollout of 5G in Europe, even when discussed outside the context of COVID-19. Importantly, in prior work, it was found that individuals who believe conspiracy theories have personality characteristics aligned with the emotions that were most strongly identified in our tweets. For example, research has found that individuals who subscribe to conspiracy theories tend to be suspicious of others, uncertain, and anxious [57].

We used dynamic topic modeling to find evidence of conspiracy theory evolution over time and to identify overlaps between theories. In Gates-classified tweets, early terms such as *predicted*, *kill_65m*, *event*, and *simulation* all refer to the simulation of a novel zoonotic coronavirus outbreak at Event 201, a global pandemic exercise that was co-hosted by several organizations, including the Bill & Melinda Gates Foundation [58]. The simulation predicted that a disease outbreak would spread to multiple countries and result in 65 million deaths. However, in April, the high-importance words shifted to include *fauci* and *buttar*, which corresponded to news coverage in which Dr Rashid Buttar stated that SARS-CoV-2 was manufactured to hurt the economy and that Dr Anthony Fauci and Bill Gates were using the pandemic to drive hidden agendas [59].

Similar morphing in the second topic of the Gates theory shows a shift in focus from funding the virus to vaccine-averse theories. Early terms such as *pirbright* and *patent* correspond to theories that the Gates Foundation funded or patented the virus through the Pirbright Institute, a UK-based company. Later, this topic morphed to include several words associated with vaccine hesitancy, such as *kennedy*, *jr*, and *fauci*, corresponding to claims by Robert Kennedy Jr that a COVID-19 vaccine would personally benefit Dr Anthony Fauci or Bill Gates. This shift in words from focusing on SARS-CoV-2 as a manufactured virus to vaccine-averse conspiracy theories highlights the importance of real-world events. Bill Gates participated in an “Ask Me Anything” on Reddit in March 2020, which highlighted Gates-funded research to develop injectable invisible ink that could be used to record vaccinations [21,22]. Immediately after this event, the prominence of words associated with vaccine-averse conspiracy theories increased, with tweets suggesting that the COVID-19 vaccine would be used to secretly microchip individuals for population control [20].

Finally, we assessed connections between conspiracy theories. Connections were most frequently identified in the Gates theory, for which nearly 13% of tweets classified as “COVID-19 misinformation” were identified in one or more of the other tracked theories. This was consistent with identified conspiracy theories connecting the Bill & Melinda Gates Foundation to work in disease research and vaccination technology. Although the Gates, Vax, and Lab theories had demonstrable overlap, only approximately 1.5% of tweets associated with 5G were found to have overlap. This may be due to the previously noted controversies surrounding the rollout of 5G in Europe.

Frequent overlap with conspiracies unrelated to COVID-19 was also observed. The Lab category showed an overlap with prior conspiracy theories about other disease outbreaks. For example, the word *ebola* was highly weighted, corresponding to the 2014-2016 Ebola outbreak, which also sparked conspiracy theories around its bioengineering or laboratory origins [11]. Other unrelated conspiracy theories were noted, including terms related to Jeffrey Epstein and his death. These observations are consistent with prior studies that showed that people who believe in one conspiracy theory are more likely to also believe in others or are more broadly prone to conspiratorial thinking [60,61].

### *Can Our Findings Inform Public Health Messaging to Reduce the Effects of Misinformation Found on Social Media?*

In exploring these four conspiracy theories, we found a clear distinction between the 5G theory and the other conspiracy theories. The 5G theory was specific and narrow in scope, while the other conspiracy theories were substantially broader, could include numerous variations on the precise actor, location, or perceived threat, and had more overlap with the other conspiracy theories overall.

It is likely that the clear scope of the 5G theory contributed to its exceptionally high classification metrics. Additionally, these distinctions in the context of public health are valuable for contextualizing any public health messaging efforts that seek to address misinformation. When determining whether to address a spreading conspiracy theory, the degree to which an emerging theory becomes entwined with existing information should determine whether the conspiracy theory should be addressed with targeted messaging versus more generalized public health information. For instance, attempts to debunk the isolated 5G connection theory were seen to elevate the exposure of the theory to a wider audience [18], while messaging regarding vaccine development and safety could both inform the public more generally and address several conspiracy theories simultaneously without promoting any particular theory.

We additionally show that conspiracy theories evolve over time by changing in focus and scope. This theory evolution will likely necessitate public health messaging, which also evolves to address a changing landscape. Our work demonstrates that off-the-shelf methods can be combined to track conspiracy theories, both in the moment and through time, to provide public health professionals with better insight into when and how to address health-related conspiracy theories. These same methods can also track public reaction to messaging to assess its impact.

## Limitations

A major limitation of any work on misinformation is that we obviously cannot examine all relevant theories, or even all of the nuance in our four identified public health-related theories, in any single study. Conspiracy theories are continuous in nature, as demonstrated here, whereas we can only observe a discrete sample within any single study. Because of this, we must aim for internal validity within any single, well-defined study and hope that many such studies will contribute to a "big picture" of social media misinformation and its effects. Not only has COVID-19 misinformation continued to spread past the end of our analysis in May 2020, but emerging conspiracy theories and topics continue to relate back to the conspiracy theories presented here. For instance, our research into claims about a laboratory origin of SARS-CoV-2 focused on popular conspiracy theories around a Chinese laboratory in Wuhan, a Canadian laboratory, and Fort Detrick in the United States. However, even at the time of this writing, two additional theories have gained traction. One indicates that the virus originated from the French Pasteur Institute; another suggests that it originated in a laboratory at the University of North Carolina [20]. We hope that results captured at the time of this analysis can inform subsequent investigations.

Second, our labeled training data explicitly labeled attempts to correct or refute misinformation as misinformation. Although this approach more accurately captured the exposure a given conspiracy might have in social media, it likely led to overestimation of the number of individuals supporting any particular theory. Excluding corrections could also have produced subtly different sentiment and dynamic topic model results, as people promoting conspiracy theories will likely differ in sentiment and word usage from those attempting to refute them. We chose to include corrections to avoid attempting to infer tweet context (eg, sarcasm is difficult to distinguish in an individual tweet) and because retweeting inaccurate information, even to correct it, still increases the number of individuals who see inaccurate content [18]. Prior work has identified both rumor-correcting and rumor-promoting tweets during crises using Twitter data [62]. Future work would benefit from considering these separately.

Additionally, our exclusive use of Twitter data fails to capture the entirety of the spread of misinformation. Social media platforms have broadly faced significant challenges in identifying and containing the spread of misinformation throughout the course of the COVID-19 pandemic [7]. Twitter users are also known to be a demographically biased sample of the US population [63-65]. Future research would benefit from analysis of misinformation on other social media platforms. Our findings are thus not generalizable to the US population as a whole. However, we emphasize that the goal of this study is not to achieve generalizability but rather to achieve internal validity by accurately categorizing sentiment and describing misinformation patterns within this population.

## Conclusions

Characterizing misinformation that poses concerns to public health is a necessary first step to developing methods to combat it. The ability to assess conspiracy theories before they become widespread would enable public health professionals to craft effective messaging to preempt misperceptions rather than to react to established false beliefs. Health officials too often fail to craft effective messaging campaigns because they target what they want to promote rather than addressing the recipients' existing misperceptions [66]. Misinformation can spread rapidly and without clear direction; this is evidenced by one tweet we uncovered while conducting this research, which shared an article promoting a conspiracy theory with the commentary that the user had not established credibility but rather "thought I'd share first" (tweet anonymized for privacy). An understanding of the appearance, transmission, and evolution of COVID-19 conspiracy theories can enable public health officials to better craft outreach messaging and to adjust those messages if public perceptions measurably shift. This study demonstrates that identifying and characterizing common and long-lived COVID-related conspiracy theories using Twitter data is possible, even when those messages shift in content and tone over time.

## Acknowledgments

We thank NewsGuard for licensing the data. Research support was provided by the UC Office of the President through its UC National Laboratory Fees Research Program (award LFR-18-547591), the Laboratory Directed Research and Development Program of Los Alamos National Laboratory (project 20200721ER), and the US Department of Energy through the Los Alamos National Laboratory. Los Alamos National Laboratory is operated by Triad National Security, LLC, for the National Nuclear Security Administration of the US Department of Energy (contract 89233218CNA000001). The Los Alamos National Laboratoory report number for this document is LA-UR-20-28305.

## Authors' Contributions

DG, NP, ARD, CWR, GF, and NYVC collected and analyzed the data. DG and ARD labeled the data for supervised learning classifiers. NP ran the dynamic topic models, which NP, DG, CDS and ARD analyzed. TP performed geospatial analyses. DG ran the sentiment analysis and built the supervised models. CDS and DG wrote the initial manuscript. All authors contributed critical revisions of the manuscript. ARD led the project.

## Conflicts of Interest

None declared.

## Multimedia Appendix 1

Supplementary figures.
[DOCX File , 2876 KB-Multimedia Appendix 1]

## References

1. Undiagnosed pneumonia - China (Hubei): request for information. ProMED International Society for Infectious Diseases. 2019 Dec 30. URL: https://promedmail.org/promed-post/?id=6864153 [accessed 2021-04-08]
2. #China has reported to WHO a cluster of #pneumonia cases —with no deaths— in Wuhan, Hubei Province. Investigations are underway to identify the cause of this illness. @WHO. 2020 Jan 04. URL: https://twitter.com/who/status/1213523866703814656?lang=en [accessed 2021-04-08]
3. According to the latest information received and @WHO analysis, there is evidence of limited human-to-human transmission of #nCOV. This is in line with experience with other respiratory illnesses and in particular with other coronavirus outbreaks. @WHOWPRO. 2020 Jan 18. URL: https://twitter.com/whowpro/status/1218741294291308545?lang=en [accessed 2021-04-08]
4. Coronavirus bioweapon – how China stole coronavirus from Canada and weaponized it. GreatGameIndia. Archived at the Internet Archive. 2020 Jan 26. URL: https://web.archive.org/web/20200313192627/https:/greatgameindia.com/coronavirus-bioweapon/ [accessed 2021-04-08]
5. Durden T. Did China steal coronavirus from Canada and weaponize it. ZeroHedge. Archived at archive.today. 2020 Jan 25. URL: http://archive.is/1EZxt#selection-803.0-824.0 [accessed 2021-04-08]
6. Deutch G. How one particular coronavirus myth went viral. Wired. 2020 Mar 19. URL: https://www.wired.com/story/opinion-how-one-particular-coronavirus-myth-went-viral/ [accessed 2020-08-18]
7. Kouzy R, Abi Jaoude J, Kraitem A, El Alam MB, Karam B, Adib E, et al. Coronavirus goes viral: quantifying the COVID-19 misinformation epidemic on Twitter. Cureus 2020 Mar 13;12(3):e7255 [FREE Full text] [doi: 10.7759/cureus.7255] [Medline: 32292669]
8. Knapp A. The original pandemic: unmasking the eerily familiar conspiracy theories behind the Russian flu of 1889. Forbes. 2020. URL: https://www.forbes.com/sites/alexknapp/2020/05/15/the-original-plandemic-unmasking-the-eerily-parallel-conspiracy-theories-behind-the-russian-flu-of-1889/#3c4e54ff50d5 [accessed 2021-04-06]
9. Ognyanova K, Lazer D, Robertson RE, Wilson C. Misinformation in action: Fake news exposure is linked to lower trust in media, higher trust in government when your side is in power. HKS Misinfo Review 2020 Jun 2;1(4):1-19. [doi: 10.37016/mr-2020-024]
10. Uscinski JE. Conspiracy Theories: A Primer. Lanham, MD: Rowman & Littlefield; 2020.
11. Turse N. How this pastor of a megachurch is fueling Ebola conspiracy theories. Time. 2019 Oct 18. URL: https://time.com/5703662/ebola-conspiracy-theories-congo/ [accessed 2020-09-09]
12. Shultz RH, Godson R. Dezinformatsia: Active Measures in Soviet Strategy. London, UK: Pergamon-Brassey's; 1984:A.
13. Jowett GS, O'Donnell V. Propaganda and Persuasion, Seventh Edition. Thousand Oaks, CA: SAGE Publications, Inc; Apr 1988:223.
14. Gorman S, Gorman J. Denying to the Grave: Why We Ignore the Facts That Will Save Us. Oxford, UK: Oxford University Press; 2016.
15. Oswald E, de Looper C. Verizon 5G: everything you need to know. Digital Trends. 2020 May 28. URL: https://www.digitaltrends.com/mobile/verizon-5g-rollout/ [accessed 2020-09-09]
16. Holmes A. 5G cell service is coming. Who decides where it goes? New York Times. 2018 Mar 02. URL: https://www.nytimes.com/2018/03/02/technology/5g-cellular-service.html [accessed 2020-09-09]
17. Morgan A. What is the truth behind the 5G coronavirus conspiracy theory? Euronews. 2020. URL: https://www.euronews.com/2020/05/15/what-is-the-truth-behind-the-5g-coronavirus-conspiracy-theory-culture-clash [accessed 2020-06-22]
18. Ahmed W, Vidal-Alaball J, Downing J, López Seguí F. COVID-19 and the 5G conspiracy theory: social network analysis of Twitter data. J Med Internet Res 2020 May 06;22(5):e19458 [FREE Full text] [doi: 10.2196/19458] [Medline: 32352383]
19. Wakefield J. How Bill Gates became the voodoo doll of Covid conspiracies. BBC News. 2020 Jun 06. URL: https://www.bbc.com/news/technology-52833706 [accessed 2020-09-09]
20. Gregory J, McDonald K. Trail of deceit: the most popular COVID-19 myths and how they emerged. NewsGuard. 2020 Jun. URL: https://www.newsguardtech.com/covid-19-myths/ [accessed 2020-09-02]
21. Goodman J, Carmichael F. Coronavirus: Bill Gates 'microchip' conspiracy theory and other vaccine claims fact-checked. BBC. 2020 May 03. URL: https://www.bbc.com/news/52847648 [accessed 2020-09-02]
22. Weintraub K. Invisible ink could reveal whether kids have been vaccinated. Scientific American. 2019 Dec 18. URL: https://www.scientificamerican.com/article/invisible-ink-could-reveal-whether-kids-have-been-vaccinated/ [accessed 2020-09-02]

23. Blancou P, Vartanian J, Christopherson C, Chenciner N, Basilico C, Kwok S, et al. Polio vaccine samples not linked to AIDS. Nature 2001 Apr 26;410(6832):1045-1046. [doi: 10.1038/35074171] [Medline: 11323657]
24. Feuer A. The Ebola conspiracy theories. New York Times. 2014 Oct 18. URL: https://www.nytimes.com/2014/10/19/sunday-review/the-ebola-conspiracy-theories.html [accessed 2020-09-02]
25. Pradhan P, Pandey A, Mishra A. Uncanny similarity of unique inserts in the 2019-NCoV spike protein to HIV-1 Gp120 and Gag. Withdrawn in: BioArxiv. Preprint posted online February 02, 2020. [doi: 10.1101/2020.01.30.927871]
26. Samorodnitsky D. Don't believe the conspiracy theories you hear about coronavirus and HIV. Especially if you work for the New York Times. Massive Science. 2020 Jan 31. URL: https://massivesci.com/notes/wuhan-coronavirus-ncov-sars-mers-hiv-human-immunodeficiency-virus/ [accessed 2020-08-07]
27. Singh M, Davidson H, Borger J. Trump claims to have evidence coronavirus started in Chinese lab but offers no details. The Guardian. 2020 Apr 30. URL: https://www.theguardian.com/us-news/2020/apr/30/donald-trump-coronavirus-chinese-lab-claim [accessed 2020-09-02]
28. Wallbank D, Bloomberg. Twitter applies another fact check—this time to China spokesman's tweets about virus origins. Fortune. 2020 May 28. URL: https://fortune.com/2020/05/28/twitter-fact-check-zhao-lijian-coronavirus-origin/ [accessed 2021-04-08]
29. Twitter flags China spokesman's tweet on COVID-19. Reuters. 2020 May 28. URL: https://www.reuters.com/article/us-twitter-china-factcheck/twitter-flags-china-spokesmans-tweet-on-covid-19-idUSKBN23506I [accessed 2021-04-08]
30. Klofstad CA, Uscinski JE, Connolly JM, West JP. What drives people to believe in Zika conspiracy theories? Palgrave Commun 2019 Apr 2;5(1). [doi: 10.1057/s41599-019-0243-8]
31. Wang Y, McKee M, Torbica A, Stuckler D. Systematic literature review on the spread of health-related misinformation on social media. Soc Sci Med 2019 Nov;240:112552 [FREE Full text] [doi: 10.1016/j.socscimed.2019.112552] [Medline: 31561111]
32. Pulido CM, Villarejo-Carballido B, Redondo-Sama G, Gómez A. COVID-19 infodemic: more retweets for science-based information on coronavirus than for false information. Int Sociol 2020 Apr 15;35(4):377-392. [doi: 10.1177/0268580920914755]
33. Broniatowski D, Kerchner D, Farooq F, Huang X, Jamison AM, Dredze M, et al. The COVID-19 social media infodemic reflects uncertainty and state-sponsored propaganda. ArXiv. Preprint posted online on July 19, 2020 [FREE Full text]
34. Memon SA, Carley KM. Characterizing COVID-19 misinformation communities using a novel Twitter dataset. ArXiv. Preprint posted online on August 3, 2020 [FREE Full text]
35. Cuan-Baltazar JY, Muñoz-Perez MJ, Robledo-Vega C, Pérez-Zepeda MF, Soto-Vega E. Misinformation of COVID-19 on the internet: infodemiology study. JMIR Public Health Surveill 2020 Apr 09;6(2):e18444 [FREE Full text] [doi: 10.2196/18444] [Medline: 32250960]
36. Gupta L, Gasparyan AY, Misra DP, Agarwal V, Zimba O, Yessirkepov M. Information and misinformation on COVID-19: a cross-sectional survey study. J Korean Med Sci 2020 Jul 13;35(27):e256 [FREE Full text] [doi: 10.3346/jkms.2020.35.e256] [Medline: 32657090]
37. Singh L, Bansal S, Bode L, Budak C, Chi G, Kawintiranon K, et al. A first look at COVID-19 information and misinformation sharing on Twitter. ArXiv. Preprint posted online on March 31, 2020 [FREE Full text]
38. Uscinski J, Enders A, Klofstad C, Seelig M, Funchion J, Everett C, et al. Why do people believe COVID-19 conspiracy theories? HKS Misinfo Review 2020 Apr 28. [doi: 10.37016/mr-2020-015]
39. Roozenbeek J, Schneider CR, Dryhurst S, Kerr J, Freeman ALJ, Recchia G, et al. Susceptibility to misinformation about COVID-19 around the world. R Soc Open Sci 2020 Oct 14;7(10):201199 [FREE Full text] [doi: 10.1098/rsos.201199] [Medline: 33204475]
40. Chaffin J. Epstein's death proves feeding ground for conspiracy theories. Financial Times. 2019 Nov 22. URL: https://www.ft.com/content/8f406516-0c9e-11ea-b2d6-9bf4d1957a67 [accessed 2021-04-08]
41. Chen E, Lerman K, Ferrara E. Tracking social media discourse about the COVID-19 pandemic: development of a public coronavirus Twitter data set. JMIR Public Health Surveill 2020 May 29;6(2):e19273 [FREE Full text] [doi: 10.2196/19273] [Medline: 32427106]
42. The internet trust tool. NewsGuard. URL: https://www.newsguardtech.com/ [accessed 2020-09-02]
43. ElSherief M, Kulkarni V, Nguyen D, Wang WY, Belding E. Hate lingo: a target-based linguistic analysis of hate speech in social media. In: Proceedings of the Twelfth International AAAI Conference on Web and Social Media (ICWSM 2018). 2018 Presented at: Twelfth International AAAI Conference on Web and Social Media (ICWSM 2018); June 25-28, 2018; Palo Alto, CA p. 42-51 URL: https://www.aaai.org/ocs/index.php/ICWSM/ICWSM18/paper/viewFile/17910/16995
44. Tay Y, Tuan L, Hui S. COUPLENET: paying attention to couples with coupled attention for relationship recommendation. In: Proceedings of the Twelfth International AAAI Conference on Web and Social Media (ICWSM 2018). 2018 Presented at: Twelfth International AAAI Conference on Web and Social Media (ICWSM 2018); June 25-28, 2018; Palo Alto, CA p. 415-424 URL: https://aaai.org/ocs/index.php/ICWSM/ICWSM18/paper/view/17828/17033
45. Daughton AR, Paul MJ. Identifying protective health behaviors on Twitter: observational study of travel advisories and Zika virus. J Med Internet Res 2019 May 13;21(5):e13090 [FREE Full text] [doi: 10.2196/13090] [Medline: 31094347]

46. Saha K, Weber I, De Choudhury M. A social media based examination of the effects of counseling recommendations after student deaths on college campuses. In: Proceedings of the Twelfth International AAAI Conference on Web and Social Media (ICWSM 2018). 2018 Presented at: Twelfth International AAAI Conference on Web and Social Media (ICWSM 2018); June 25-28, 2018; Palo Alto, CA p. 320-329 URL: https://aaai.org/ocs/index.php/ICWSM/ICWSM18/paper/view/17855/17023
47. Rizoiu MA, Graham T, Zhang R, Zhang Y, Ackland R, Xie L. DEBATENIGHT: the role and influence of socialbots on Twitter during the first 2016 U.S. Presidential Debate. In: Proceedings of the Twelfth International AAAI Conference on Web and Social Media (ICWSM 2018). 2018 Presented at: Twelfth International AAAI Conference on Web and Social Media (ICWSM 2018); June 25-28, 2018; Palo Alto, CA p. 300-309 URL: https://aaai.org/ocs/index.php/ICWSM/ICWSM18/paper/view/17886/17021
48. AFINN. 2011 Mar. URL: http://www2.imm.dtu.dk/pubdb/pubs/6010-full.html [accessed 2021-04-08]
49. Mohammad SM. NRC Word-Emotion Association Lexicon (Aka EmoLex). Saif M Mohammad. 2010. URL: http://saifmohammad.com/WebPages/NRC-Emotion-Lexicon.htm [accessed 2020-09-01]
50. Blei D, Lafferty J. Dynamic topic models. In: ICML '06: Proceedings of the 23rd international Conference on Machine learning. 2006 Jun Presented at: 23rd International Conference on Machine Learning; June 25-29, 2006; Pittsburgh, PA p. 113-120.
51. Blei D, Ng A, Jordan M. Latent Dirichlet allocation. J Mach Learn Res 2003;3:993-1022 [FREE Full text]
52. Řehůřek R, Sojka P. Software framework for topic modelling with large corpora. In: Proceedings of LREC 2010 Workshop New Challenges for NLP Frameworks. 2010 Presented at: LREC 2010 Workshop New Challenges for NLP Frameworks; May 22, 2010; Valetta, Malta p. 45-50.
53. Gerrish S, Blei D. A language-based approach to measuring scholarly impact. In: ICML'10: Proceedings of the 27th International Conference on International Conference on Machine Learning. 2010 Presented at: 27th International Conference on International Conference on Machine Learning; June 21-24, 2010; Haifa, Israel p. 375-382.
54. Ferré-Sadurní L, McKinley J. Alex Jones is told to stop selling sham anti-coronavirus toothpaste. New York Times. 2020 Mar 13. URL: https://www.nytimes.com/2020/03/13/nyregion/alex-jones-coronavirus-cure.html [accessed 2020-09-01]
55. Stanway D. China lab rejects COVID-19 conspiracy claims, but virus origins still a mystery. Reuters. 2020 Apr 28. URL: https://www.reuters.com/article/us-health-coronavirus-china-lab/china-lab-rejects-covid-19-conspiracy-claims-but-virus-origins-still-a-mystery-idUSKCN22A0MM [accessed 2020-09-02]
56. Chan AKM, Nickson CP, Rudolph JW, Lee A, Joynt GM. Social media for rapid knowledge dissemination: early experience from the COVID-19 pandemic. Anaesthesia 2020 Dec;75(12):1579-1582 [FREE Full text] [doi: 10.1111/anae.15057] [Medline: 32227594]
57. Goreis A, Voracek M. A systematic review and meta-analysis of psychological research on conspiracy beliefs: field characteristics, measurement instruments, and associations with personality traits. Front Psychol 2019 Feb 11;10:205 [FREE Full text] [doi: 10.3389/fpsyg.2019.00205] [Medline: 30853921]
58. The Event 201 scenario. Event 201. URL: https://www.centerforhealthsecurity.org/event201/scenario.html [accessed 2020-09-01]
59. Ohlheiser A. How covid-19 conspiracy theorists are exploiting YouTube culture. MIT Technology Review. 2020 May 7. URL: https://www.technologyreview.com/2020/05/07/1001252/youtube-covid-conspiracy-theories/ [accessed 2020-09-01]
60. Lewandowsky S, Cook J. The Conpsiracy Theory Handbook. George Mason University Center for Climate Change Communication. 2020 Mar. URL: https://www.climatechangecommunication.org/wp-content/uploads/2020/03/ConspiracyTheoryHandbook.pdf [accessed 2021-04-08]
61. Swami V, Voracek M, Stieger S, Tran US, Furnham A. Analytic thinking reduces belief in conspiracy theories. Cognition 2014 Dec;133(3):572-585. [doi: 10.1016/j.cognition.2014.08.006] [Medline: 25217762]
62. Arif A, Robinson J, Stanek S. A Closer Look at the Self-Correcting Crowd: Examining Corrections in Online Rumors. In: CSCW '17: Proceedings of the 2017 ACM Conference on Computer Supported Cooperative Work and Social Computing. 2017 Feb Presented at: 2017 ACM Conference on Computer Supported Cooperative Work and Social Computing; February 2017; Portland, OR p. 155-168. [doi: 10.1145/2998181.2998294]
63. Chou WS, Hunt YM, Beckjord EB, Moser RP, Hesse BW. Social media use in the United States: implications for health communication. J Med Internet Res 2009 Nov 27;11(4):e48 [FREE Full text] [doi: 10.2196/jmir.1249] [Medline: 19945947]
64. Mislove A, Lehmann S, Ahn Y, Onnela J, Rosenquist J. Understanding the demographics of Twitter users. In: Proceedings of the Fifth International AAAI Conference on Weblogs and Social Media. 2011 Presented at: Fifth International AAAI Conference on Weblogs and Social Media; July 17-21, 2011; Barcelona, Spain p. 554-557 URL: https://www.aaai.org/ocs/index.php/ICWSM/ICWSM11/paper/viewFile/2816/3234
65. Duggan M, Ellison NB, v C, Lenhart A, Madden M. Demographics of Key Social Networking Platforms. Pew Research Center. 2015 Jan 09. URL: http://www.pewinternet.org/2015/01/09/demographics-of-key-social-networking-platforms-2/ [accessed 2016-08-02]
66. Larson HJ. The biggest pandemic risk? Viral misinformation. Nature 2018 Oct 16;562(7727):309-309. [doi: 10.1038/d41586-018-07034-4] [Medline: 30327527]

## Abbreviations

**API:** application programming interface
**DTM:** dynamic topic modeling
**NRC:** National Research Council
**WHO:** World Health Organization